\documentclass[a4paper, amsfonts, amssymb, amsmath, reprint, showkeys, nofootinbib, twoside]{revtex4-1}
\usepackage{graphicx,color}
\usepackage[utf8]{inputenc}
   \usepackage{amsmath,amssymb,pifont}

\newcommand{\Ds}{\displaystyle}

\newcommand{\nn}{\nonumber}

\newcommand{\Tr}{\mathrm{Tr}}

\newcommand{\ot}{\leftarrow}

\renewcommand{\(}{\left(}
\renewcommand{\)}{\right)}
\renewcommand{\[}{\left[}
\renewcommand{\]}{\right]}

\newcommand{\fnot}[1]{\not{\! #1}}

\begin{document}
\title{Transverse-momentum-dependent factorization for lattice observables}
\author{Alexey A. Vladimirov and Andreas Sch\"afer}
\affiliation{Institut f\"ur Theoretische Physik, Universit\"at Regensburg,\\
D-93040 Regensburg, Germany}

\begin{abstract}
Using soft collinear effective field theory, we derive the factorization theorem for the quasi-transverse-momentum-dependent (quasi-TMD) operator. We check the factorization theorem at one-loop level and compute the corresponding coefficient function and anomalous dimensions. The factorized expression is built from the physical TMD distribution, and a nonperturbative lattice related factor. We demonstrate that lattice related functions cancel in appropriately constructed ratios. These ratios could be used to explore various properties of TMD distributions, for instance, the nonperturbative evolution kernel. A discussion of such ratios and the related continuum properties of TMDs is presented.
\end{abstract}
\maketitle

\section{Introduction}

Over the last years, continuous progress in theory and phenomenology of a transverse-momentum-dependent (TMD) factorization theorem made it a valuable tool for analysis and prediction of many observables (for a review see \cite{Angeles-Martinez:2015sea}). It has been demonstrated that the TMD factorization approach accurately describes the data in a broad range of energies and a wide spectrum of processes \cite{Bacchetta:2017gcc,Scimemi:2017etj,Vladimirov:2019bfa,Scimemi:2019cmh,Bacchetta:2019sam}. Conceptually, TMD factorization \cite{Collins:2011zzd,GarciaEchevarria:2011rb} is different from the collinear factorization and gives rise to a number of specific novel effects. In this article, we apply TMD factorization to a certain class of operators suitable for evaluation by QCD lattice methods.

There are eight leading twist TMD distributions, each of which depends on a transverse variable and longitudinal momentum fraction $x$. The purely experimental determination of all of these TMD distributions is a highly nontrivial task. Therefore, the prospects for obtaining complementary information from QCD lattice simulations look extremely promising, in particular, due to the possibility of measuring correlators directly in coordinate space. The latter point is advantageous because the TMD factorization theorem we discuss and TMD distributions are naturally formulated in coordinate space, despite the fact that their interpretation is usually given in momentum space. From experimental data, one can extract coordinate space information only via a Fourier transformation, resulting in a significant systematic error and model-bias. A good example for the encountered problems is the TMD evolution kernel $\mathcal{D}(b)$ (also known as Collins-Soper(CS) kernel \cite{Collins:1981va}). To extract the CS kernel from data, one has to combine data from many experiments performed at varying energies. The current global pool of data gives access to energies from 1 to 150~GeV \cite{Scimemi:2019cmh}. However, the precision of most of these data is quite limited, and their interpretation depends nontrivially on $\mathcal{D}$. The later is known up to $\alpha_s^3$-order in perturbation theory \cite{Vladimirov:2016dll}, but is poorly constrained beyond perturbative values of $b$. Even the shape of $\mathcal{D}(b)$ is questionable (compare for instance the extractions made in \cite{Scimemi:2019cmh} and \cite{Bacchetta:2019sam}, and their discussion in  Ref.\cite{Vladimirov:2020umg}). This problem can be resolved, or at least reduced by lattice simulations.

Suggestions for lattice studies of TMD observables were made long ago \cite{Musch:2007ya,Musch:2010ka}. At that time, however, some crucial assumptions were rather conjectural. Recently, such efforts were promoted to a higher level with the formulation of appropriate factorization theorems \cite{Ebert:2018gzl,Ebert:2019okf,Ji:2019ewn}. In all cases, one considers an equal-time analog of a TMD operator, which turns into an ordinary TMD operator after the boost. In this paper, we present a different analysis of the same operator within the TMD factorization approach, based on the $q_T$-dependent soft-collinear effective field theory (SCET II). We demonstrate that the suggested lattice observables are more closely related to TMD hadron tensor rather than to the TMD distributions. Using the TMD hadron tensor, we present a construction that stresses the analogy between lattice observables and physical quantities used in the description of conventional processes like the Drell-Yan process, utilizing the same terminology.  Although the route of the derivation of the factorized expression differs from that of \cite{Ebert:2019okf,Ji:2019ewn}, we arrive at an equivalent result. Checking the factorized expression at the one-loop level, we found that the perturbative parts are in complete agreement with \cite{Ebert:2019okf}. 

The expression derived here is only the leading term of the factorization theorem. The subsequent terms are formally suppressed by powers of hadron momentum. This fact should not be over interpreted because a closer analysis reveals the potential breakdown of this expansion. Namely, each next order term has a stronger small-$x$ singularity than the preceding ones. Such a problem is quite common for factorization theorems for lattice observables. For example, the quasiparton distribution functions (PDF) \cite{Ji:2013dva}, and pseudo-PDFs \cite{Radyushkin:2017cyf} also suffer from this problem, as is discussed, e.g., in Ref.\cite{Braun:2018brg}.  In the TMD case, the small-$x$ divergent terms are more troublesome because they are enhanced by TMD evolution. This is unavoidable since the hard scale is the parton momentum $\sim x P$ leading to strong factorization breaking at small $x$.

The paper is split into three sections. In Sec.\ref{sec:factorization} we define matrix elements suitable for lattice simulations and derive the factorized expressions using SCET II. In Sec.\ref{sec:NLO} we check factorization at the one-loop level and derive the corresponding coefficient function at the next-to-leading order (NLO). Finally, in Sec.\ref{sec:ratios} we discuss ratios of matrix elements that have simpler properties and thus could serve to determine TMD distributions with less effort. We emphasize, in particular, the advantages of ratios at small-longitudinal separation.

\section{Factorization theorem}
\label{sec:factorization}

\subsection{Definition of lattice observables.} 

The considered lattice observables read
\begin{eqnarray}\label{def:Sigma}
&&W^{[\Gamma]}_{f\ot h}(b;\ell,L;v,P,S)=\frac{1}{2}\langle P,S|\bar q_f(b+\ell v)\Gamma 
\\ \nn &&\qquad \qquad\times [b+\ell v,b+L v][b+Lv,Lv][Lv,0] q_f(0)|P,S\rangle,
\end{eqnarray}
where $|P,S\rangle$ is a single-hadron state with momentum $P$ and spin $S$, and $\Gamma$ is a Dirac matrix. The hadron species and the flavor of the quark field are specified by the labels $h$ and $f$. $[x,y]$ is the straight gauge link between points $x$ and $y$,
\begin{eqnarray}
[x,y]=P\exp\(ig\!\int_{0}^1 \!\!dt (x-y)^\mu A_\mu(y+t(x-y))\).
\end{eqnarray}
The same object has been considered in \cite{Zhang:2018ggy,Ebert:2019okf,Ebert:2018gzl,Ebert:2019tvc,Ji:2019ewn}. Often, the matrix element (\ref{def:Sigma}) is called a quasi-TMD distribution in analogy to the quasiparton distribution functions \cite{Ji:2013dva}. However, as we demonstrate in the next section, the structure of (\ref{def:Sigma}) does not remind of a TMD distribution but rather of a TMD hadronic tensor, such as the hadronic tensor for Drell-Yan or Semi-Inclusive Deep Inelastic Scattering (SIDIS). For that reason, we avoid the term quasi-TMD distribution, and denote (\ref{def:Sigma}) by the letter $W$.

\begin{figure}[t]
\includegraphics[width=0.4\textwidth]{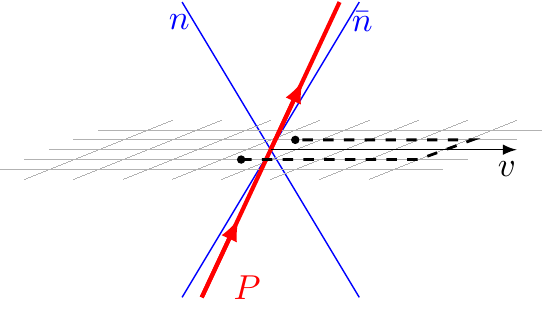}
\includegraphics[width=0.3\textwidth]{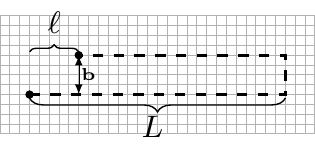}
\caption{\label{fig:lattice-operator} Illustration for the definition of the matrix element $W(b;\ell,L;v,P)$ (\ref{def:Sigma}). Dashed lines denotes the Wilson links, and black dots denote the quark fields. Top and bottom illustrations correspond to side and top views relative to the gauge link plane.}
\end{figure}

The space-time orientation of different quantities in $W$ is given in Fig.\ref{fig:lattice-operator}. The lattice operator must be an equal-time operator and thus the vectors $v^\mu$ and $b^\mu$ do not have time components. Consequently, $b^2<0$ and $v^2<0$. The vectors $P^\mu$ and $v^\mu$ define an analog of the scattering plane. The transverse vector $b^\mu$ is orthogonal to them,
\begin{eqnarray}\label{def:vb=bP=0}
(vb)=0,\qquad (bP)=0.
\end{eqnarray}
The vector $b^\mu$ defined by (\ref{def:vb=bP=0}) is restricted to a line, due to the absence of time components (except for the  special case that the vectors $v^\mu$, $P^\mu$ and the time direction lie in a plane). The situation is different for physical kinematics where the scattering plane is formed by two timelike vectors and thus the vector $b^\mu$ is restricted to a plane. 

\subsection{Factorization limit} 

The clear separation of collinear and soft-field modes within the hadron is a prerequisite for any TMD factorization theorem. It can be achieved by considering a fast moving hadron, for which anticollinear components of field momenta are suppressed in comparison to collinear ones. To quantify this condition we write the momentum of a hadron as
\begin{eqnarray}
P^\mu=P^+ \bar n^\mu+\frac{M^2}{2P^+}n^\mu,
\end{eqnarray}
where $M$ is the mass of the hadron,  $n$ and $\bar n$ are lightlike vectors $n^2=\bar n^2=0$ (see also Fig.\ref{fig:lattice-operator}), normalized according to $(n\bar n)=1$.  Here, we use the standard notation of light-cone components of a vector $a^\mu$:
\begin{eqnarray}
a^\mu=\bar n^\mu a^++n^\mu a^-+a^\mu_T.
\end{eqnarray}
So, the factorization limit requires
\begin{eqnarray}\label{condition1}
\frac{P^-}{P^+}=\frac{M^2}{2(P^+)^2}\sim \lambda^2 \ll 1.
\end{eqnarray}
with $\lambda$ being the generic small parameter of SCET. In this regime the hadron momentum is almost lightlike. 

We also assume that the staple-shaped gauge links contour (\ref{def:Sigma}) is much longer than broad
\begin{eqnarray}\label{condition2}
b,\ell\ll L.
\end{eqnarray} 
Under this assumption, the effects caused by the interaction with the transverse gauge link $[b+Lv,Lv]$ are suppressed as $b/L$ and $\ell/L$, and thus can be neglected. We then introduce an ``instant'' (formally infinitely heavy)  scalar field, $H(x)$, with the Lagrangian
\begin{eqnarray}\label{L_HH}
\mathcal{L}_{HH}=H^\dagger (ivD)H+\mathcal{O}\(L^{-1}\),
\end{eqnarray}
and approximate the gauge links $[x,x+L v]$ by the $H$ propagator. In eq.(\ref{L_HH}), $D_\mu$ is the QCD covariant derivative. The field $H$ differs from a usual scalar heavy quark field \cite{Eichten:1989zv} only by the fact that $v^2<0$. (This is why we call it ``instant'' field.)

In the notation (\ref{L_HH}), the similarity of the matrix element (\ref{def:Sigma}) with the ordinary hadron tensor for TMD factorization becomes transparent. We rewrite (\ref{def:Sigma}) as
\begin{eqnarray}\label{hadron-tensor}
&&W_{ij}(b;\ell,L;v,P)=
\\&& \qquad\nn 
\sum_{X}\langle P,S|J_i^\dagger(v\ell+b)|X\rangle \langle X|J_j(0)|P,S\rangle+\mathcal{O}\(L^{-1}\),
\end{eqnarray}
where $W^{[\Gamma]}=\frac{1}{2}\Tr(W\Gamma)$ and $J_i$ is the heavy-to-light current
\begin{eqnarray}\label{def:current}
J_i(x)=H^\dagger(x)q_i(x).
\end{eqnarray}
The structure of the first term in (\ref{hadron-tensor}) resembles the structure of the hadron tensor for TMD cross sections, with the main difference that there is only a single hadron. The second hadron is replaced by the ``instant'' field $H$.

\subsection{Field modes factorization and SCET current.} 

\begin{figure}[t]
\includegraphics[width=0.45\textwidth]{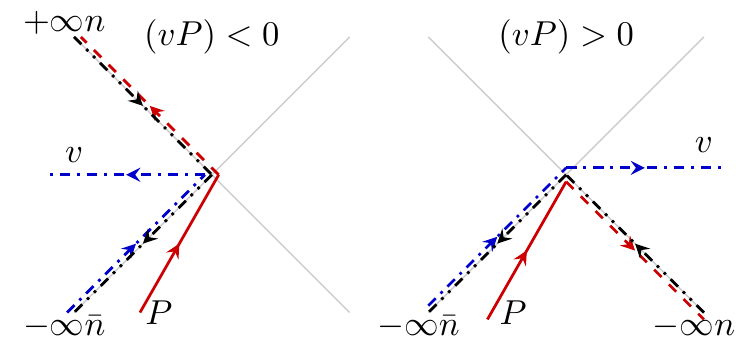}
\caption{\label{fig:infinities} Illustration for Wilson lines structure for operators $J_i$ (\ref{def:SCET-J}) and $\overline{J}_i$ (\ref{def:SCET-J:pv<0}) in the plane $(P,v)$. Red (solid and dashed) color indicates the collinear fields, blue (dot-dashed) -- anticollinear fields and black (double-dot-dashed) -- soft fields.}
\end{figure}

The analogy with the TMD hadron tensor (\ref{hadron-tensor}) allows us to recapitulate the main points of TMD factorization and apply it to the lattice case. To derive the factorized expression, we use the soft-collinear effective field theory with finite $q_T$ (SCET II) approach, similarly to Ref.\cite{GarciaEchevarria:2011rb}.

In SCET II, one distinguishes collinear, anticollinear and soft fields. In leading approximation, the fast-moving hadron is a composition of collinear fields ($\xi$ for quarks and $A_{c,\mu}$ for gluons). Their momentum components are of the structure
\begin{eqnarray}\label{counting}
\partial_\mu \xi \sim \{1,\lambda^2,\lambda\}\xi, \qquad \partial_\mu A_c\sim\{1,\lambda^2,\lambda\}A_c.
\end{eqnarray}
The separation of transverse and collinear modes requires a clear hierarchy between corresponding momentum components. In the present case, the typical transverse momenta in diagrams are $\sim b^{-1}$. And, therefore, we have an additional constraint
\begin{eqnarray}
\frac{1}{|b|P^+}\sim \lambda.
\end{eqnarray}

\textit{A priory} it is not evident how to count the field $H$ in terms of $\lambda$ since a Wilson line does not carry a momentum. However, the situation becomes clear if one boosts the system such that $P^-\to 0$. Then the Wilson line $H$ is turned towards the light-cone direction $n$. In the boosted frame, the $v^+$-component of $v^\mu$ can be ignored and the field $H$ can be approximately considered as an anticollinear Wilson line. Therefore, the fields $H$ and collinear fields have no direct interaction but only through soft exchanges.

Using these counting rules we write the leading power SCET operator that corresponds to the current $J_i$ (\ref{def:current}) as
\begin{eqnarray}\label{def:SCET-J}
J_i^{\text{SCET}}(x)=C_H(v\hat p)(H^\dagger W_{\bar n})Y^\dagger_{\bar n} Y_{n}(W_{n}^\dagger \xi_i)(x).
\end{eqnarray}
In this expression the collinear Wilson line
\begin{eqnarray}\label{def:collinearWL}
W_{n}(x)=P\exp\(ig \int_{-\infty}^0d\sigma n A_{c}(x+n \sigma)\),
\end{eqnarray}
contains all gluons radiated by the collinear quark field. The anticollinear Wilson line $W_{\bar n}$ contains the gluons radiated by the $H$-field and is given by a similar expression with $n\to \bar n$. The Wilson lines $Y$ are the result of the decoupling transformation \cite{Bauer:2001yt}. They have analogous expression to (\ref{def:collinearWL}) but build with soft gluon fields.

The coefficient $C_H$ is the matching coefficient between SCET and QCD operators. It depends on the momentum of field $\xi$ (in position space $\hat p$ is an operator), and is independent of quark flavor. At leading order $C_H=1$.

Expression (\ref{def:SCET-J}) applies for $(vP)>0$. If $(vP)<0$ the gluon fields are summed with the opposite sign (in comparison to (\ref{def:SCET-J})) and thus they form Wilson lines pointing to $+\infty n$. So, for the case $(vP)<0$ the SCET operator reads
\begin{eqnarray}\label{def:SCET-J:pv<0}
\overline{J}_i^{\text{SCET}}(x)=\overline{C}_H(v\hat p)(H^\dagger W_{\bar n})Y^\dagger_{\bar n} \overline{Y}_{n}(\overline{W}_{n}^\dagger \xi_i)(x),
\end{eqnarray}
where
\begin{eqnarray}\label{def:collinearWL-}
\overline{W}_{n}(x)=P\exp\(ig \int_{+\infty}^0d\sigma n A_{c}(x+n \sigma)\),
\end{eqnarray}
and similarly for $\overline{Y}$. Alternatively, the directions of Wilson lines could be recovered by noting that in the $P^-\to0$ boosted frame, Wilson lines $H$ point to the past (future) if $(vP)>0$ ($(vP)<0$). A visual representation of operator $J$ and $\overline{J}$ is shown in Fig.\ref{fig:infinities}.

Combining (\ref{def:SCET-J}) and (\ref{def:SCET-J:pv<0}) we express the QCD current (\ref{def:current}) by
\begin{eqnarray}\label{J-to-SCET}
J_i(x)=\left\{\begin{array}{cc}
J_i^{\text{SCET}}(x)+\mathcal{O}(\lambda), & ~~~{\rm for}~~~(vP)>0,
\\
\overline{J}_i^{\text{SCET}}(x)+\mathcal{O}(\lambda), & ~~~{\rm for}~~~(vP)<0.
\end{array}\right.
\end{eqnarray}
Note, that different spinor components of $\xi_i$ have different power counting. The large (small) components can be projected out by the matrix $\gamma^-\gamma^+/2$ ($\gamma^+\gamma^-/2$).

\subsection{Factorized expression.}

Substituting the effective currents (\ref{J-to-SCET}) into the expression (\ref{def:Sigma}) we obtain (here for $(vP)>0$)
\begin{widetext}
\begin{eqnarray}\label{factorization0}
&&W_{f\ot h}^{[\Gamma]}(b;\ell,L;v,P,S)=
\\&&\nn\qquad
\Big|C_H(\hat p v)\Big|^2 \langle P,S| \(\bar \xi W_{n}(b+\ell v)\frac{\Gamma}{2} W^\dagger_{n}\xi(0)\)
\(H^\dagger W_{\bar n}(0)W^\dagger_{\bar n}H(b+\ell v)\) \frac{\text{Tr}}{N_c}\[Y^\dagger_{ n} Y_{\bar n}(b+\ell v)Y^\dagger_{\bar n} Y_{n}(0)\] | P,S\rangle.
\end{eqnarray}
Here, we have performed a Fiertz transformation to recouple color indices, and have dropped color-covariant structures. The collinear,  anticollinear and soft fields operate on different Hilbert spaces, such that the total Hilbert space can be written as a direct product of three distinct Hilbert spaces \cite{Bauer:2000yr,Lee:2006nr}. Doing so one has to correct for overlap of the field definitions in the soft region. The overlap contribution can be removed by the so-called zero-bin subtraction factor \cite{Manohar:2006nz}. Additionally the fields can be Taylor-expanded in the slow (in comparison to other components) directions, which are determined by the counting rules (\ref{counting}). After these operations we obtain the following result
\begin{eqnarray}\label{factorization1}
&&W_{f\ot h}^{[\Gamma]}(b;\ell,L;v,P,S)=
\Big|C_H(\hat p v)\Big|^2\widetilde \Phi_{f\ot h}^{[\Gamma']}(b,\ell v^-;P,S)
\widetilde \Psi(b,\ell v^+;v) \frac{S(b)}{\text{Z.b.}}
+\mathcal{O}\(\lambda\),
\end{eqnarray}
where
\begin{eqnarray}\label{def:Gamma'}
\Gamma'=\frac{1}{4}\gamma^+\gamma^-\Gamma\gamma^-\gamma^+.
\end{eqnarray}
The functions are 
\begin{eqnarray}
\widetilde \Phi_{f\ot h}^{[\Gamma]}(b,x^-;P,S)&=&\langle P,S|\bar q(x^- n+b)[x^-  n+b,-\infty n+ b]\frac{\Gamma}{2}
[-\infty  n,0]q(0)|P,S\rangle,
\\\label{def:Psi}
\widetilde \Psi(b,x^+;v)&=&\langle 0|H^\dagger(0) [0,-\infty \bar n][-\infty \bar n+b,x^+\bar n+b]H(x^+\bar n+b)|0\rangle,
\\
S(b)&=&\frac{\text{Tr}}{N_c}\langle 0|[b,-n\infty +b][-n\infty,0][0,-\bar n\infty][-\bar n\infty +b,b] |0\rangle,
\end{eqnarray}
\end{widetext}
where we use the QCD fields since within its own Hilbert space each sector of SCET is equivalent to QCD. The zero-bin factor (denoted as Z.b.) removes the contribution from the overlap of soft and collinear modes. It is not known explicitly except for certain regularizations, for instance for $\delta$-regularization \cite{GarciaEchevarria:2011rb} where it is equivalent to the TMD soft factor.

The function $\widetilde \Phi$ is (a Fourier transform of) an unsubtracted TMD distribution. The function $\widetilde{\Psi}$ has an analogous structure. The only difference is that it measures a TMD distribution of the field $H$. For that reason we call it an (unsubtracted) instant-jet TMD distribution. The function $S$ is the TMD soft factor. The expression (\ref{factorization1}) applies for $(vP)>0$ and thus, $\widetilde \Phi$ and $S$ correspond to Drell-Yan kinematics. If $(vP)<0$ the Wilson line along $n$ points to $+\infty n$, and $\widetilde \Phi$ and $S$ correspond to SIDIS kinematics (and the coefficient function is replaced by $\overline{C}_H$).

\subsection{Recombination of rapidity divergences.} 

Unsubtracted TMD distributions have rapidity divergences that appear due to the presence of infinite lightlike Wilson lines separated in the transverse plane. Rapidity divergences are associated with the directions of Wilson lines. In the current case, there are two light-cone directions, and thus we introduce two regularization parameters $\delta^+$ and $\delta^-$. These parameters regularize rapidity divergences associated with the directions $n$ and $\bar n$, correspondingly. In the product of all functions in (\ref{factorization1}) rapidity divergences cancel, and therefore, the last step of the factorization approach is to recombine rapidity divergences, and to introduce physical (aka finite) TMD distributions. In this procedure we follow Ref.\cite{Vladimirov:2017ksc}. In what follows, we use $\delta$-regularization for rapidity divergences, but the same procedure can be performed for different kinds of regulators. The final result is independent of the regularization used.

In the expression (\ref{factorization1}) the rapidity divergences are present according to the following pattern (in this section we omit all arguments of the different functions, except the ones related to rapidity divergences):
\begin{eqnarray}\label{rap.div.1}
W=|C_H|^2\widetilde{\Phi}(\delta^+)\frac{S(\delta^+,\delta^-)}{\text{Z.b.}(\delta^+,\delta^-)}\widetilde{\Psi}(\delta^-).
\end{eqnarray}
In Ref.\cite{Vladimirov:2017ksc} it has been shown that rapidity divergences are structurally equivalent to ultraviolet divergences, and therefore, can be absorbed into a divergent factor $R$. Introducing rapidity renormalization factors into (\ref{rap.div.1}) we obtain
\begin{eqnarray}\label{rap.div.2}
W=|C_H|^2\widetilde{\Phi}(\nu^+)S^{-1}_0(\nu^2)\widetilde{\Psi}(\nu^-),
\end{eqnarray}
where $\nu^\pm$ are the scales of rapidity-divergence renormalization, and 
\begin{eqnarray}\nn
&&\widetilde{\Phi}(\nu^+)=\widetilde{\Phi}(\delta^+)R\(\frac{\delta^+}{\nu^+}\),
\quad
\widetilde{\Psi}(\nu^-)=R\(\frac{\delta^-}{\nu^-}\)\widetilde{\Psi}(\delta^-),
\\\nn 
&&S_0^{-1}(\nu^2)=R^{-1}\(\frac{\delta^+}{\nu^+}\)\frac{S(\delta^+,\delta^-)}{\text{Z.b}(\delta^+,\delta^-)}R^{-1}\(\frac{\delta^-}{\nu^-}\).
\end{eqnarray}
The function $S_0$ depends on $\nu^2=2\nu^+\nu^-$ due to Lorenz invariance. This expression is independent of $\nu^\pm$ by definition, and each function here is finite. The dependence on $\nu^\pm$ is given by the renormalization group equation
\begin{eqnarray}\label{dnuPhi}
\nu^+\frac{d \widetilde{\Phi}(\nu^+)}{d\nu^+}=\frac{\mathcal{D}}{2}\widetilde{\Phi}(\nu^+),
\end{eqnarray}
where $\mathcal{D}$ is the rapidity anomalous dimension \cite{Vladimirov:2017ksc}, or CS-kernel \cite{Collins:1981va}:
\begin{eqnarray}
\mathcal{D}=\frac{1}{2}\frac{d \ln R}{d\ln\nu^+}.
\end{eqnarray}
 The equation for $\widetilde{\Psi}$ is analogous.

Introducing the boost-invariant variables
\begin{eqnarray}\label{def:zetas}
\zeta=2(P^+)^2\frac{\nu^-}{\nu^+},\qquad \bar \zeta=2\mu^2(v^-)^2\frac{\nu^+}{\nu^-},
\end{eqnarray}
where $P^+$ is the collinear component of hadron momentum and $\mu$ is a factorization scale. The parameter $\zeta$ is the standard rapidity evolution parameter \cite{GarciaEchevarria:2011rb,Vladimirov:2017ksc,Collins:2011zzd}. The parameter $\bar\zeta$ is the analogous parameter for $\Psi$. Let us emphasize that the expression for $\bar \zeta$ is unusual, because typically the scale of rapidity divergences is associated with the collinear component of a momentum. However, there is no momentum that is associated with the field $H$. The only momentum scale presented in $\Psi$ is the factorization scale. In Sec.\ref{sec:NLO}, we confirm (\ref{def:zetas}) by a one-loop calculation. The dependence of the function $\widetilde{\Phi}(\zeta,\nu^2)$ on $\zeta$ follows from (\ref{dnuPhi}) and reads
\begin{eqnarray}
\zeta\frac{d \widetilde{\Phi}(\zeta,\nu^2)}{d\zeta}=-\mathcal{D}\widetilde{\Phi}(\zeta,\nu^2).
\end{eqnarray}
The function $\Psi$ depends on $\bar \zeta$ in the same way.

Generally, the function $S_0(\nu^2,b)$ is a process-dependent and (at large-$b$) nonperturbative function. To get rid of it, we note that the variable $\nu^2$ decouples from the evolution, and thus the function $S_0$ can simply be absorbed into the definition of a TMD distribution.
In fact, the physical definition of TMD distributions already includes such factors (see discussion in \cite{Vladimirov:2017ksc,Collins:2011zzd}). They are built from the remnants of TMD soft factors. So, a physical TMD distribution, such as the one used to describe the Drell-Yan or SIDIS processes, is defined together with an appropriate $S_0^{\text{TMD}}$  as
\begin{eqnarray}\label{Phi->physical}
\Phi(\zeta)=\frac{\widetilde{\Phi}(\zeta,\nu^2)}{\sqrt{S_0^{\text{TMD}}(\nu^2)}}.
\end{eqnarray}
It is independent on $\nu^2$~\cite{Vladimirov:2017ksc}. To formulate the factorization in terms of physical TMD distributions, we use the definition (\ref{Phi->physical}) and compensate the extra factor $\sqrt{S^{\text{TMD}}_0}$ by an appropriate redefinition of the instant-jet TMD distribution:
\begin{eqnarray}\label{Psi->physical}
\Psi(\bar\zeta)=\frac{\sqrt{S_0^{\text{TMD}}(\nu^2)}\widetilde{\Psi}(\bar\zeta,\nu^2)}{S_0(\nu^2)}.
\end{eqnarray}
Note, that in a suitably defined regularization scheme (for instance $\delta$-regularization \cite{GarciaEchevarria:2011rb}), the zero-bin subtraction factor $\text{Z.b}=S^2(b)$, and thus $S_0^{\text{TMD}}=S_0$. However, generally, these factors can be different in the nonperturbative regime.

\subsection{Final form of the factorized expression.}

The final form of the factorized expression reads
\begin{eqnarray}\label{factorization2}
&&W_{f\ot h}^{[\Gamma]}(b;\ell,L;v,P,S;\mu)=
\\\nn &&\Big|C_H\(\frac{\hat p v}{\mu}\)\Big|^2\Phi_{f\ot h}^{[\Gamma']}(b,\ell v^-;\mu,\zeta;P,S)
\Psi(b,\ell v^+;\mu,\bar \zeta;v)
\\\nn &&\qquad\qquad\qquad\qquad\qquad
+\mathcal{O}\(\frac{P^-}{P^+},\frac{1}{|b|P^+},\frac{b}{L},\frac{\ell}{L}\).
\end{eqnarray}
Here, we restored all arguments of functions including the scale $\mu$. 

There are two points in equation (\ref{factorization2}) which need clarification. The first point concerns the dependence of $\Psi$ on the variable $\ell$. The variable $\ell$ appears in $\Psi$ accompanied by the lightlike vector $\bar n$, and thus can enter the function only in a scalar product with some other vector of the problem. This can only be the vector $v$, and thus the dependence on $\ell$ can appear only as $(v^+v^-\ell \Lambda_{\text{QCD}})$ or as $(v^+v^-\ell/L)$ in the presence of a regularization parameter $L$  (compare to the function $\Phi$ where the vector $\ell$ enters via $(\ell v^- P^+)$). However, in the factorization limit both of these combinations are negligible. Thus, we conclude that the dependence on $\ell$ is marginal,
\begin{eqnarray}\label{Psi-on-l}
\Psi(b,\ell v^+;\mu,\bar \zeta;v)=\Psi(b;\mu,\bar \zeta)+\mathcal{O}\(\frac{\ell}{L},\ell\Lambda_{\text{QCD}}\).
\end{eqnarray}
This statement is also clear from the boosted frame perspective: boosting $P^-/P^+\to\infty$ one automatically gets $v^+/v^-\to0$. So, the dependence on  $\ell$ is negligible, unless $\ell$ is very large. Note, that the function $\Psi$ defined in (\ref{Psi->physical}) is not an equal-time observable, and thus could not be computed on the lattice directly.

The second point concerns the definition of the operator $\hat p=-i\partial_\ell$, acting on the function $\Phi$. To rewrite it in an explicit form we recall the definition of TMD distributions as functions of the momentum fraction
\begin{eqnarray}
&&\Phi_{f\ot h}^{[\Gamma ]}(x,b;\mu,\zeta)
=
\\\nn&& \qquad\int \frac{dy^-}{2\pi}e^{-ixy^-P^+}\Phi_{f\ot h}^{[\Gamma]}(b,y^-;\mu,\zeta;P,S).
\end{eqnarray}
In this representation, the positive and negative values of $x$ are related to quark and antiquark distributions
\begin{eqnarray}
&&\Phi^{[\Gamma ]}_{f\ot h}(x,b;\mu,\zeta)=
\\\nn &&\qquad\theta(x)\Phi^{[\Gamma ]}_{f\ot h}(x,b;\mu,\zeta)-(-1)^r \theta(-x)\Phi^{[\Gamma ]}_{\bar f\ot h}(|x|,b;\mu,\zeta),
\end{eqnarray}
where $r$ depends on $\Gamma$ and the Lorentz structure of the TMD. For example, $\Gamma=\gamma^+$, $r=0(1)$ for unpolarized (spin-flip) TMDs.  

Using these facts we rewrite (\ref{factorization2}) as
\begin{eqnarray}\label{factorization3}
&&W^{[\Gamma]}_{f\ot h}(b;\ell,L;v,P,S;\mu)=
\\\nn &&\frac{1}{P_v}\int dx e^{ix\ell P_v}\Big|C_H\(\frac{|x| P_v}{\mu}\)\Big|^2\Phi_{f\ot h}^{[\Gamma']}(x,b;\mu,\zeta)
\Psi(b;\mu,\bar \zeta)
\\\nn &&\qquad\qquad\qquad\qquad
+\mathcal{O}\(\frac{P^-}{P^+},\frac{1}{|b|P^+},\frac{b}{L},\frac{\ell}{L},\ell\Lambda_{\text{QCD}}\),
\end{eqnarray}
where $P_v=v^-P^+$. This is probably the most practical form of the factorization theorem, and we will use it later. The factorization statement is independent on the Dirac structure, which is standard for the TMD factorization approach. Therefore, using this expression one can describe polarized and unpolarized processes equally well.

It is important to emphasize that the size of power corrections in (\ref{factorization3}) significantly depends on $x$. In fact, the typical momentum scale entering factorized expressions is $\hat p\sim x P$ rather then just $P$. Therefore, a more reliable estimation of the power corrections is $\mathcal{O}(P^-/x^2P^+)$. This is a typical size of power corrections to factorization theorems for lattice observables, see e.g. the case of quasi-PDF power correction which are of order  $\mathcal{O}(\Lambda^2/x^2(pv)^2)$ as is shown in Ref.\cite{Braun:2018brg}. Such a large power correction can undermine the applicability of the whole approach as we discuss below.

\section{NLO expressions}
\label{sec:NLO}

In this section we present the computation of elements of the factorization theorem at one loop. The calculation supports the correctness of the construction. The calculation presented here is done in the $\delta$-regularization scheme \cite{GarciaEchevarria:2011rb,Echevarria:2016scs}, which allows us to reuse results of earlier calculations made in \cite{Echevarria:2016scs,Echevarria:2015byo}. Our results coincide with those of \cite{Ebert:2018gzl}, where they were reached in a different manner.

\subsection{Hard matching coefficient}

\begin{figure}[b]
\includegraphics[width=0.3\textwidth]{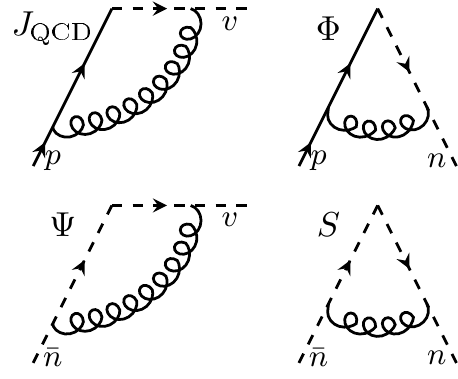}
\caption{\label{fig:diag-hard} Diagrams to be computed for evaluation of the hard matching coefficient. Solid (dashed) lines represent the quark field (Wilson line). In the case $(vP)>0$ ($<0$) the Wilson lines along $n$ point to $-\infty$ ($+\infty$).}
\end{figure}

To evaluate the matching coefficient between the QCD current (\ref{def:current}) and the SCET current (\ref{def:SCET-J}) one needs to compute and compare both sides of equations (\ref{J-to-SCET}). At the same time, one should demonstrate cancellation of collinear and soft divergences. We use $\delta$-regularization for collinear and soft divergences and dimensional regularization ($d=4-2\epsilon$) for ultraviolet divergences ($\epsilon>0$).

In the $\delta$-regularization scheme \cite{GarciaEchevarria:2011rb,Echevarria:2016scs} the zero-bin subtraction factor coincides with the soft factor squared
\begin{eqnarray}
\text{Z.b}|_{\delta-\text{reg.}}=S^2(b).
\end{eqnarray}
Therefore, the matching relation at NLO turns into
\begin{eqnarray}\label{NLO:matching}
C_H^{[1]}J_{\text{QCD}}^{[0]}=J_{\text{QCD}}^{[1]}-\Phi^{[1]}-\Phi^{[0]}\Psi^{[1]}+\Phi^{[0]}S^{[1]},
\end{eqnarray}
where we omit arguments for simplicity, and use the following shorthand notation for coefficients of perturbative series $X=X^{[0]}+\alpha_s X^{[1]}+...$~. We have also used that $\Psi^{[0]}=S^{[0]}=1$. There are four diagrams that contribute to (\ref{NLO:matching}), presented in Fig.\ref{fig:diag-hard}. There exist other diagrams contributing to each term of (\ref{NLO:matching}), (these are various self-energy diagrams),
but their contributions exactly cancel in the sum. 

In $\delta$-regularization the diagram $J_{\text{QCD}}$ reads
\begin{eqnarray}\nn
J^{[1]}_{\text{QCD}}=\int\frac{d^dk}{(2\pi)^d}\frac{-ig^2C_F(\fnot k+\fnot p)\fnot v u_{\bar n}}{[(p+k)^2+i\Delta][k^2+i0][(kv)+i\Delta_v]},
\end{eqnarray}
where $\Delta$ and $\Delta_v$ are parameters of $\delta$-regularization ($\Delta>\Delta_v>0$). Evaluating this diagram in the limit $\Delta$, $\Delta_v$ $\to0$ and $p^+\gg p^-$ we obtain
\begin{eqnarray}\label{nlo:J}
&& J^{[1]}_{\text{QCD}}=u_{\bar n}\frac{\alpha_s}{2\pi}C_F e^{-i\epsilon \pi}\Big\{
\\\nn && -(-(pv)-i0)^{\epsilon}(i\Delta)^{-\epsilon}(i\Delta_v)^{-\epsilon}\Gamma^2(\epsilon)\Gamma(1-\epsilon)
\\\nn && -(-v^2)^{\epsilon}(2i\Delta_v)^{-2\epsilon}\Gamma(2\epsilon)\Gamma(-\epsilon)
+(i\Delta)^{-\epsilon}\frac{\Gamma(\epsilon)}{\epsilon(1+\epsilon)}
\\\nn && +(-v^2)^{\epsilon}(-2(pv)- i0)^{-2\epsilon}\frac{\Gamma(-1+2\epsilon)\Gamma(2-\epsilon)}{\epsilon}\Big\}+...~,
\end{eqnarray}
where the dots stand for power suppressed contributions $\sim \Delta$. The $i0$-terms are important for proper analytic continuation between the cases $(pv)>0$ and $(pv)<0$. It reads $(-(pv)- i0)=|pv|e^{i(\text{arg}(pv)-\pi)}$. The first term in brackets represents the soft divergence, whereas the second and the third term are collinear and anticollinear divergences. Note, that $\epsilon>0$ throughout and thus factors $\Delta^{-\epsilon}$ are divergent at $\Delta\to0$. 

Evaluating analogously the rest of the diagrams (note, that the results for $\Phi^{[1]}$ and $S^{[1]}$ in  $\delta$-regularization can be found in refs.\cite{GarciaEchevarria:2011rb,Echevarria:2016scs} and \cite{GarciaEchevarria:2011rb,Echevarria:2015byo}, correspondingly) we obtain
\begin{eqnarray}\nn
\Phi^{[1]}&=&u_{\bar n}\frac{\alpha_s}{2\pi}C_F e^{-i\epsilon \pi}\Big\{(i\Delta)^{-\epsilon}\frac{\Gamma(\epsilon)}{\epsilon(1-\epsilon)}
\\ &&
-(\mp (pn))^\epsilon (i\Delta)^{-\epsilon}(i\delta^+)^{-\epsilon}\Gamma^2(\epsilon)\Gamma(1-\epsilon)\Big\},
\\\nn
\Psi^{[1]}&=&\frac{\alpha_s}{2\pi}C_F e^{-i\epsilon \pi}\Big\{-(-v^2)^\epsilon(2i\Delta_v)^{-2\epsilon}\Gamma(2\epsilon)\Gamma(-\epsilon)
\\ &&\label{nlo:Psi}
-(-(v\bar n))^\epsilon (2i\Delta_v)^{-\epsilon}(i\delta^-)^{-\epsilon}\Gamma^2(\epsilon)\Gamma(1-\epsilon)\Big\},
\\\label{nlo:S}
S^{[1]}&=&-\frac{\alpha_s}{2\pi}C_F e^{-i\epsilon \pi}(\pm 2\delta^+\delta^-)^{-\epsilon} \Gamma^2(\epsilon)\Gamma(1-\epsilon),
\end{eqnarray}
where the upper sign corresponds to the geometrical configuration with $(vP)>0$ and the lower sign corresponds to configuration with $(vP)<0$. The regularized propagators reproduce soft propagators in the soft regime, therefore, the parameters $\Delta$ and $\Delta_v$ are related to $\delta^\pm$ according to 
\begin{eqnarray}
\delta^-=\frac{\Delta}{2p^+},\qquad \delta^+=\frac{\Delta_v}{\pm v^-}.
\end{eqnarray}
The sign of $v^-$ is the same as the sign of $(pv)$, and thus $\delta^+>0$. Substituting these expressions into (\ref{NLO:matching}) we observe that each divergent sector cancels exactly (i.e. at all orders of $\epsilon$-expansion). Clearly, it is important to keep the proper direction of Wilson lines in mind, which leads to different signs of $(vP)$ resulting in these cancellations. Altogether, this confirms the derived factorization theorem at NLO. 

The hard-matching coefficient is
\begin{eqnarray}
&&C_H(pv)=1+C_F\frac{\alpha_s}{2\pi}e^{i\epsilon (\pi-2\text{arg}(pv))}
\\\nn&&\qquad \times(2|pv|)^{-2\epsilon}\frac{\Gamma(-1+2\epsilon)\Gamma(2-\epsilon)}{\epsilon}
+\mathcal{O}(\alpha_s^2).
\end{eqnarray}
Performing renormalization in the $\overline{\text{MS}}$-scheme we obtain
\begin{eqnarray}\label{CH2}
&&\Big|C_H\(\frac{vp}{\mu}\)\Big|^2=1+
\\&&\nn\qquad
C_F\frac{\alpha_s}{4\pi}\(-\mathbf{L}^2+2\mathbf{L}-4+\frac{\pi^2}{6}\)+\mathcal{O}(\alpha_s),
\end{eqnarray}
where $\mathbf{L}=\ln((2|pv|)^2/\mu^2)$. Importantly, the coefficient function is the same for $(pv)>0$ and $(pv)<0$ at this perturbative order. Nonetheless, the continuation between these regions is nontrivial, and for higher orders the coefficient functions could be different. The expression (\ref{CH2}) coincides with the one derived for the hard coefficient function in Ref.\cite{Ebert:2018gzl,Ebert:2019okf}, where the calculation has been done differently.

We have also performed the same computation with a finite-length $H$-Wilson line, as in (\ref{def:Sigma}). At $L\to\infty$ the results coincide with (\ref{nlo:J}-\ref{nlo:S}) after the replacement $\Delta_n^{n\epsilon}\to L^{-n\epsilon}\Gamma(1+n\epsilon)$. The cancellation of divergences also takes place, although the matching relation between $\delta^+$ and $L$ depends on $\epsilon$ and does not hold at higher orders of perturbation theory.

\subsection{Anomalous dimensions}

The functions $\Phi$ and $\Psi$ are TMD distributions and obey the double scale evolution
\begin{eqnarray}\label{evol:1}
&&\frac{d\ln\Phi^{[\Gamma]}(b,x^-;\mu,\zeta;P,S)}{d\ln\mu^2}=\frac{\gamma_F(\mu,\zeta)}{2},
\\
&&\frac{d\ln\Phi^{[\Gamma]}(b,x^-;\mu,\zeta;P,S)}{d\ln\zeta}=-\mathcal{D}(b,\mu),
\end{eqnarray}
and
\begin{eqnarray}
&&\frac{d\ln\Psi^{[\Gamma]}(b,x^+;\mu,\zeta;v)}{d\ln\mu^2}=\frac{\gamma_{\Psi}(\mu,\zeta)}{2},
\\\label{evol:4}
&&\frac{d\ln\Psi^{[\Gamma]}(b,x^+;\mu,\zeta;v)}{d\ln\zeta}=-\mathcal{D}(b,\mu),
\end{eqnarray}
where $\gamma_F$ and $\gamma_{\Psi}$ are ultraviolet anomalous dimensions, and $\mathcal{D}$ is the rapidity anomalous dimension. The integrability condition for these equations gives the Collins-Soper equation 
\begin{eqnarray}
\!\!\frac{d \gamma_F(\mu,\zeta)}{d\ln \zeta}=\frac{d \gamma_{\Psi}(\mu,\zeta)}{d\ln \zeta}=\!-\frac{d \mathcal{D}(b,\mu)}{d\ln \mu}
=\!-\Gamma_{\text{cusp}}(\mu),
\end{eqnarray}
where $\Gamma_{\text{cusp}}$ is the cusp anomalous dimension for lightlike Wilson lines. The solution for ultraviolet anomalous dimensions is
\begin{eqnarray}
\gamma_F(\mu,\zeta)&=&\Gamma_{\text{cusp}}(\mu)\ln\(\frac{\mu^2}{\zeta}\)-\gamma_V(\mu),
\\
\gamma_{\Psi}(\mu,\zeta)&=&\Gamma_{\text{cusp}}(\mu)\ln\(\frac{\mu^2}{\zeta}\)-\overline{\gamma_{\Psi}}(\mu).
\end{eqnarray}
The anomalous dimension $\gamma_V$ is known up to $\alpha_s^3$-order, and at LO $\gamma_V=-6C_F \alpha_s/(4\pi)$. The anomalous dimension $\overline{\gamma_{\Psi}}$ is the finite part of the heavy-to-light anomalous dimension. Since the vertex diagram (\ref{nlo:Psi}) contributes only to the double logarithm structure, the finite part of the anomalous dimension is twice the anomalous dimension of a heavy quark field \cite{Broadhurst:1991fz}
\begin{eqnarray}
\overline{\gamma_{\Psi}}(\mu)=8C_F\frac{\alpha_s}{4\pi}+\mathcal{O}(\alpha_s^2).
\end{eqnarray}

All components of the current $J_i$ are renormalized by a single renormalization factor, $J_i^{\text{ren.}}=Z_{J} J_{i}$, and thus the matrix element $W$ is renormalized by
\begin{eqnarray}
W^{[\Gamma]}_{\text{ren.}}=Z_{J}^2W^{[\Gamma]}.
\end{eqnarray}
The corresponding anomalous dimension
\begin{eqnarray}
\mu^2 \frac{d W^{[\Gamma']}}{d\mu^2}=\gamma_J W^{[\Gamma']},
\end{eqnarray}
is evaluated in \cite{Chetyrkin:2003vi} at NNLO (for $v^2>0$) and reads
\begin{eqnarray}
\gamma_J=-3C_F\frac{\alpha_s}{4\pi}+\mathcal{O}(\alpha_s^2).
\end{eqnarray}
Note, that the anomalous dimensions $\gamma_J$ and $\overline{\gamma_{\Psi}}$ are known from the literature and are related to heavy quarks physics (i.e. with $v^2>0$). They agree with those presented here at LO. However, they could disagree at higher perturbative orders due to $v^2<0$ kinematics.

The renormalization group requires
\begin{eqnarray}\nn
\frac{d \ln |C_H(pv/\mu)|^2}{d\ln \mu^2}+\frac{\gamma_V(\mu,\zeta)+\gamma_{\Psi}(\mu,\bar \zeta)}{2}=\gamma_J(\mu).
\end{eqnarray}
Substituting LO anomalous dimensions and the NLO coefficient function (\ref{CH2}), we check that this relation is satisfied if the $\zeta$-parameters are related by
\begin{eqnarray}
\zeta\bar \zeta=(2\hat p^+v^-)^2\mu^2.
\end{eqnarray}
This fixes the relative freedom in the definition of boost-invariant variables $\zeta$ and $\bar \zeta$. It also confirms our observation that in the absence of momentum the rapidity divergences in $\overline{\Psi}$ are weighted by the factorization scale $\mu$ (\ref{def:zetas}).

\section{Ratios of lattice observables}
\label{sec:ratios}

The factorized expression (\ref{factorization3}) has the generic form of a TMD factorization theorem, and, therefore, it incorporates three nonperturbative functions. These are the TMD distribution $\Phi$, the instant-jet TMD distribution $\Psi$, and the rapidity anomalous dimension $\mathcal{D}$. The latter is not explicitly presented in the formula, but enters via the scaling properties of the distributions, see (\ref{evol:1}-\ref{evol:4}). To determine these functions one needs to measure $W$ in a large range of parameters $P$ and $\ell$. However, even in this case the function $\Psi$, which depends on $\ell$ only weakly (\ref{Psi-on-l}), would be totally correlated with $\mathcal{D}$.

There are two principal ways to bypass this problem. The first approach is to obtain the values of $\Psi$ in an independent calculation. This could be done in perturbation theory (at small values of $b$) \cite{Ebert:2019okf}, or performing a separate lattice calculation, such as the one suggested recently in Ref.\cite{Ji:2019sxk}. The second approach is to consider ratios of lattice observables, such that undesired factors (in particular the function $\Psi$) cancel. This approach looks more promising because the measurements of ratios are simpler on the lattice. In addition to the cancellation of $\Psi$ one would also profit from the cancellation of various other multiplicative factors such as lattice renormalization constants, and a corresponding reduction of systematic uncertainties for the lattice results.

In this section, we consider ratios of the form
\begin{eqnarray}\label{def:R}
R=\frac{W^{[\Gamma_1]}_{f_1\ot h_1}(b;\ell,L;v,P_1,S_1;\mu)}{W^{[\Gamma_2]}_{f_2\ot h_2}(b;\ell,L;v,P_2,S_2;\mu)}.
\end{eqnarray}
In such ratios the contribution of $\Psi$ cancels, as well as a common $\mu$-dependence. Various properties of these ratios have been considered in \cite{Musch:2010ka,Musch:2011er,Engelhardt:2015xja,Ebert:2018gzl,Ebert:2019tvc}. In the subsequent sections, we discuss particularly interesting combinations of parameters in $R$, that were not yet mentioned in the literature. These combinations allow to check the validity of the factorization theorem, and estimate the joined systematic uncertainties of the approach and the lattice computation. Additionally, we consider the case $\ell=0$. The $\ell=0$ case is particularly simple to simulate on the lattice, as has been done in refs.\cite{Musch:2011er,Engelhardt:2015xja} for TMD double moments. We show that it grants access to the nonperturbative anomalous rapidity dimension.

For brevity of the formulas in this section we denote only those arguments of $W$ that are different, assuming that all the other arguments in the ratio (\ref{def:R}) are the same. Also, we universally denote all power corrections by
\begin{eqnarray}
\mathcal{O}(\lambda)=\mathcal{O}\(\frac{P^-}{x^2P^+},\frac{1}{|b|P^+},\frac{b}{L},\frac{\ell}{L},\ell\Lambda_{\text{QCD}}\).
\end{eqnarray}

\subsection{Sign-flip}

The most elementary test of the factorization theorem (\ref{factorization3}) is the measurement of the famous sign-flip of P-odd TMD distributions between the Drell-Yan and SIDIS process \cite{Collins:2002kn}. SIDIS and Drell-Yan kinematics are distinguished  by the sign of $(vP)$. Therefore, the sign-flip can be tested by replacing $v\to-v$.

For example, considering $\Gamma=\gamma^+$. We have two Lorentz structures
\begin{eqnarray}
W^{[\gamma^+]}=W_1+i\epsilon_T^{\mu\nu}b_\mu s_\nu W_{1T}^\perp,
\end{eqnarray}
where $\epsilon_T^{\mu\nu}=\epsilon^{+-\mu\nu}$. These structures can be independently extracted from a lattice simulation \cite{Musch:2011er}, and are proportional to unpolarized $f_1$ and Sivers $f_{1T}^\perp$ TMD distributions. The Sivers distribution is P-odd and its sign depends on the direction of the Wilson lines, in contrast to unpolarized distributions.  Therefore, the following ratios should hold
\begin{eqnarray}\label{sign-flip}
\frac{W_1(-v)}{W_1(v)}=1+\mathcal{O}(\alpha_s^2,\lambda),
\\\nn
\frac{W_{1T}^\perp(-v)}{W_{1T}^\perp(v)}=-1+\mathcal{O}(\alpha_s^2,\lambda).
\end{eqnarray}
These relations are trivial at NLO due to the independence of coefficient $|C_H|^2$ of the sign of $(vP)$. However, they could be violated by higher perturbative terms, if there are nontrivial effects of analytical continuation in $(vP)$.

Similar measurements have been performed in \cite{Musch:2011er,Engelhardt:2015xja}, where the ratios $ W_{1T}^\perp/W_1$ (and similar for $\Gamma=i\sigma^{+\mu}\gamma^5$) has been studied at different values of $b$ and $P$ and for different signs of $v$. Perfect agreement with (\ref{sign-flip}) has been demonstrated.

\subsection{Power suppressed terms}

A great feature of lattice QCD is the possibility to measure objects unaccessible in an experiment directly. In particular, one can compare measurements of different Lorentz structures and check
the TMD factorization theorem in a completely controlled environment. The Dirac structures of higher TMD twist must be suppressed due to dominance of the collinear components in the hadron. We have
\begin{eqnarray}\label{R:suppresed}
\frac{W_{f\ot h}^{[\Gamma_1]}}{W_{f\ot h}^{[\Gamma_2]}}=\mathcal{O}(\lambda),
\end{eqnarray}
where $\Gamma_1'=0$ and $\Gamma_2'=\Gamma_2$ with $\Gamma'$ defined in (\ref{def:Gamma'}).

Despite the apparent triviality of this statement the numerical evaluation of this ratio on the lattice
is very important because it allows to estimate systematic uncertainties. In a sense, it directly measures the size of power corrections to the factorization theorem (\ref{factorization3}). This is very valuable information, because the accessible hadron momenta in state-of-the-art lattice simulations are at most a few GeV, such that power corrections can be large.

\subsection{Nonperturbative rapidity anomalous dimension}

The most exciting property of the ratios $R$ is their exclusive sensitivity to the rapidity anomalous dimension, which was also pointed out in refs.\cite{Ebert:2018gzl,Ebert:2019tvc,Ji:2019ewn}. The properly constructed ratio is almost independent of nonperturbative functions except the rapidity anomalous dimension. To extract $\mathcal{D}$ one needs the ratio of $W$'s at different momenta
\begin{eqnarray}
R_{P_1/P_2}=\frac{W_{f\ot h}^{[\Gamma]}(P_1)}{W_{f\ot h}^{[\Gamma]}(P_2)}.
\end{eqnarray}
Using (\ref{factorization3}) we get
\begin{widetext}
\begin{eqnarray}
R_{P_1/P_2}=\frac{P_2^+}{P_1^+}
\frac{\Ds\int dx_1 e^{ix_1\ell v^-P_1^+}\Big|C_H\(\frac{x_1 v^-P_1^+}{\mu}\)\Big|^2 \Phi_{f\ot h}^{[\Gamma']}(x_1,b;\mu,\zeta_1)}{\Ds\int dx_2 e^{ix_2\ell v^-P_2^+}\Big|C_H\(\frac{x_2 v^-P_2^+}{\mu}\)\Big|^2 \Phi^{[\Gamma']}_{f\ot h}(x_2,b;\mu,\zeta_2)}+\mathcal{O}(\lambda),
\end{eqnarray}
where $\zeta_1=c_0(2|x_1v^-|P_1^+)^2 $, $\zeta_2=c_0(2|x_2v^-|P^+_2)^2 $ with $c_0$ being a constant. Note, that the scale $\mu$ is taken to be the same in the numerator and the denominator in order to cancel the unknown functions $\Psi$. Evolving the remaining two functions in $\zeta$ to the same point $\zeta_0$, and partially canceling evolution factors we get
\begin{eqnarray}\label{ratio1}
R_{P_1/P_2}=\(\frac{P_2^+}{P_1^+}\)^{2\mathcal{D}(b,\mu)+1}\frac{
\Ds\int dx_1 e^{ix_1\ell v^-P_1^+}\Big|C_H\(\frac{|x_1 v^-|P_1^+}{\mu}\)\Big|^2 \Phi^{[\Gamma']}_{f\ot h}(x_1,b;\mu,\zeta_0)
|x_1|^{-2\mathcal{D}(b,\mu)}}{\Ds\int dx_2 e^{ix_2\ell v^-P_2^+}\Big|C_H\(\frac{|x_2 v^-|P_2^+}{\mu}\)\Big|^2 \Phi^{[\Gamma']}_{f\ot h}(x_2,b;\mu,\zeta_0)
|x_2|^{-2\mathcal{D}(b,\mu)}}+\mathcal{O}(\lambda).
\end{eqnarray}
\end{widetext}
A similar ratio is considered in detail in Ref.\cite{Ebert:2019tvc},where it is suggested to calculate the Fourier images of denominator and numerator separately. Such a method has bright prospects, but is noticeably more demanding on the lattice side than the one discussed below. The main difficulty comes from the noncancellation of lattice renormalization factors that, therefore, have to be computed separately. An additional, but not smaller, problem comes from the Fourier transformation in $\ell$, which has to be deduced from the few  measurable points with $\ell\ll L,\Lambda_{\text{QCD}}^{-1}$. Altogether, these problems could result in a large systematic uncertainty.

To avoid these difficulties we suggest to consider the case of $\ell=0$. Roughly speaking, the plain $\ell=0$ case corresponds to the ratio of the first Mellin moments of TMD distributions. The higher moments can be accessed by taking derivatives with respect to $\ell$. Let us denote
\begin{eqnarray}\label{def:Rn}
\mathbf{R}^{(n)}=\(\frac{P_2^+}{P_1^+}\)^{n-2}\frac{\partial_\ell^{n-1} W^{[\Gamma]}_{f\ot h}(P_1)}{\partial_\ell^{n-1}  W^{[\Gamma]}_{f\ot h}(P_2)}\Bigg|_{\ell=0},
\end{eqnarray}
where the prefactor is chosen such that at $b\to0$ the ratios become unity $\mathbf{R}^{(n)}\to 1$. These ratios give direct access to the rapidity anomalous dimension, as we demonstrate below. The lattice computation of the $n=1$ case is relatively straightforward, and some exploratory computations were already made in \cite{Musch:2010ka,Musch:2011er,Engelhardt:2015xja}. The consideration of higher $n$ is more complicated due to the mixture of operators with different Dirac structures (especially, if $b$ is not much smaller than $L$, see also \cite{Shanahan:2019zcq}), and possible problems with restoration of rotational symmetries. Nonetheless, we expect that these problems could be overcome and, at least, the case $n=2$ is feasible.

\begin{figure}[t]
\includegraphics[width=0.45\textwidth]{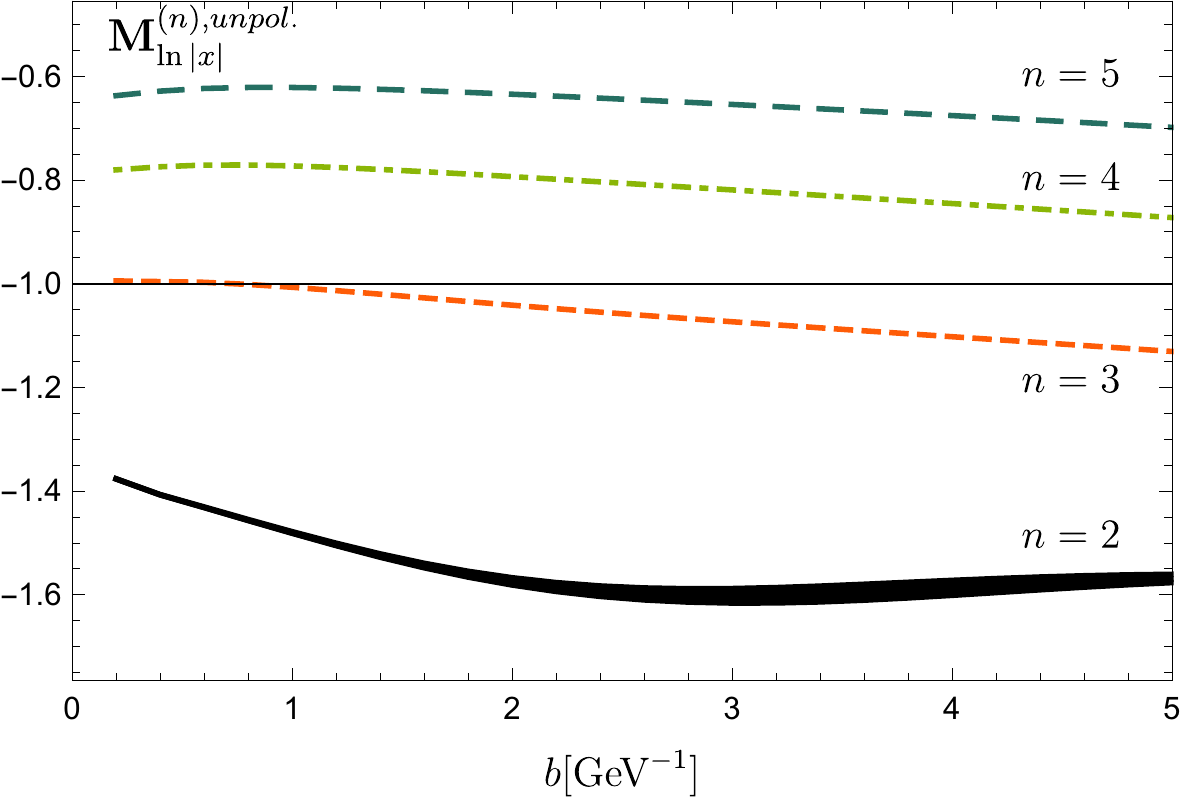}
\caption{\label{fig:Mn} Functions $\mathbf{M}^{(n),\Gamma}_{\ln|x|}$ at different values of $b$ and $n$. The computation is done for unpolarized TMD PDF (nonsinglet combination $u-d$) and $\mathcal{D}$ extracted in Ref.\cite{Scimemi:2019cmh}, at $\mu=3.2$GeV. The integral over $x$ has been cut at $|x|>x_0$, the thickness of the line demonstrates the resulting uncertainty, as obtained by the variation $x_0^{-4\pm1}$. The case $n=1$ is divergent and not presented.}
\end{figure}

The ratio (\ref{def:Rn}) at $\ell=0$ is convenient to present in the form
\begin{eqnarray}
\mathbf{R}^{(n)}=\(\frac{P_2^+}{P_1^+}\)^{2\mathcal{D}(b,\mu)}\mathbf{r}^{(n)}+\mathcal{O}(\lambda),
\end{eqnarray}
where $\mathbf{r}^{(n)}=1+\mathcal{O}(\alpha_s)$. The NLO contribution to $\mathbf{r}^{(n)}$ is obtained using (\ref{CH2})
\begin{eqnarray}\label{def:rn}
&&\mathbf{r}^{(n)}=1+4C_F\frac{\alpha_s(\mu)}{4\pi}\ln\(\frac{P^+_1}{P^+_2}\)
\Big[
\\\nn &&\qquad 1-\ln\(\frac{4P_1^+P_2^+|v^-|^2)}{\mu^2}\)-2\mathbf{M}^{(n),\Gamma}_{\ln|x|}(b,\mu)\Big]+\mathcal{O}(\alpha_s^2),
\end{eqnarray}
where
\begin{eqnarray}\label{def:M}
&&\mathbf{M}^{(n),\Gamma}_{f(x)}(b,\mu)=
\\\nn &&\qquad \frac{\int dx f(x)~|x|^{-2\mathcal{D}(b,\mu)+n-1}\Phi_{f\ot h}^{[\Gamma]}(x,b;\mu,\zeta_0)}{\int dx |x|^{-2\mathcal{D}(b,\mu)+n-1}\Phi_{f\ot h}^{[\Gamma]}(x,b;\mu,\zeta_0)}.
\end{eqnarray}
It is straightforward to check that this expression is independent on $\mu$ and $\zeta_0$. Therefore, it can be further simplified by using the optimal definition of a TMD distribution \cite{Scimemi:2018xaf}. Setting $\zeta_0=\zeta_\mu(b)$ such that $\Phi_{f\ot h}^{[\Gamma]}(x,b;\mu,\zeta_\mu(b))=\Phi_{f\ot h}^{[\Gamma]}(x,b)$ is independent on $\mu$, we obtain
\begin{eqnarray}\nn
\mathbf{M}^{(n),\Gamma}_{f(x)}(b,\mu)=\frac{\int dx f(x)~|x|^{-2\mathcal{D}(b,\mu)+n-1}\Phi_{f\ot h}^{[\Gamma]}(x,b)}{\int dx |x|^{-2\mathcal{D}(b,\mu)+n-1}\Phi_{f\ot h}^{[\Gamma]}(x,b)},
\end{eqnarray}
where $\Phi^{[\Gamma]}(x,b)$ (without scale) is the optimal TMD distribution. The optimal definition is convenient and often used for phenomenological extractions \cite{Scimemi:2018xaf,Bertone:2019nxa,Scimemi:2017etj,Scimemi:2019cmh,Vladimirov:2019bfa}. Let us mention, that the NNLO expression for $\mathbf{r}^{(n)}$ can be derived using only the NLO anomalous dimensions and the finite part of $|C_H|^2$ at NLO. Thus, it could be possibly reconstructed from already existing calculations.

\begin{figure}[t]
\includegraphics[width=0.45\textwidth]{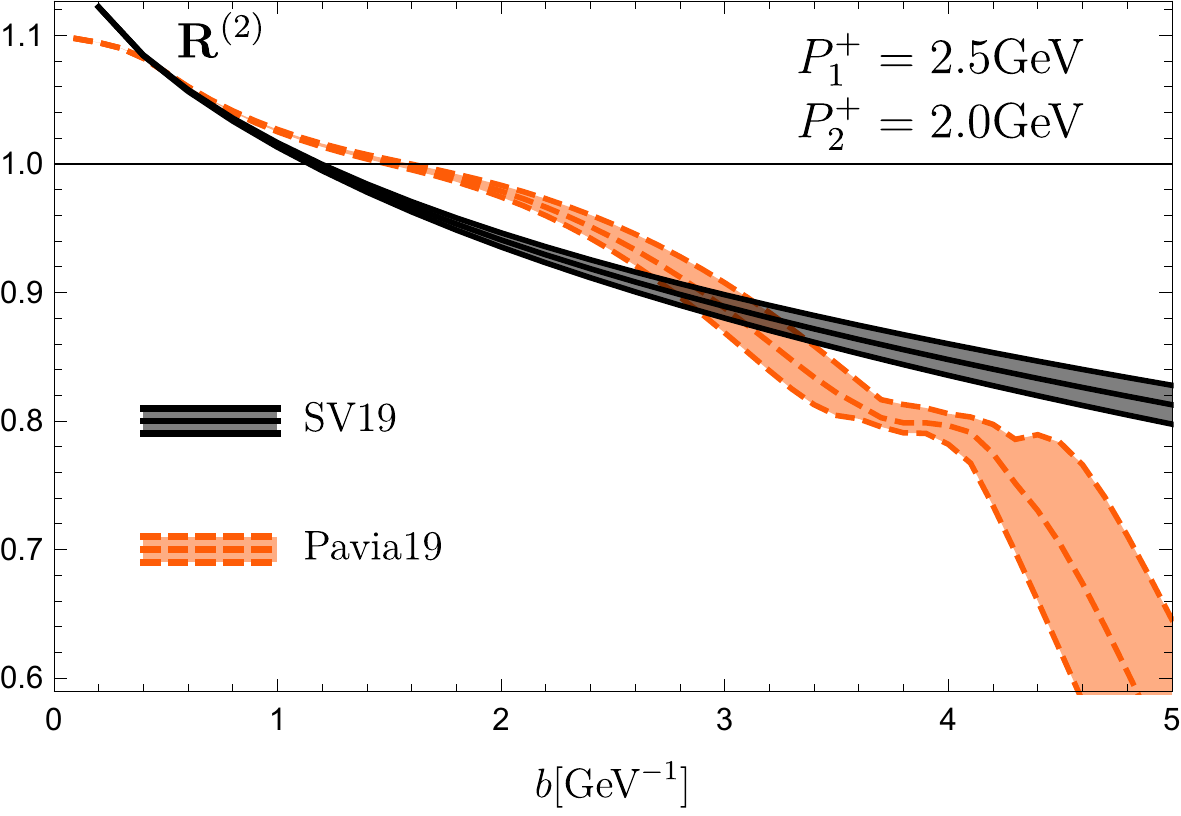}
\caption{\label{fig:compare} Comparison of the functions $\mathbf{R}^{(2)}$ evaluated with two different phenomenological models for rapidity anomalous dimension: ``SV19'' and ``Pavia19'' that are considered in \cite{Scimemi:2019cmh} and \cite{Bacchetta:2019sam}, correspondingly. The uncertainty band is obtained by variation of model parameters within their uncertainty range.}
\end{figure}

\begin{table*}
\begin{tabular}{|c|c|c||l|}
 $\Gamma$ & Kind & Name & Comment
\\\hline
$\gamma^+$ & $f_1$ &  Unpolarized & Divergent in full range of $b$.
\\\hline
$\gamma^+$ & $f_{1T}^\perp$ &  Sivers & Possibly divergent.
\\\hline
$\gamma^+\gamma^5$ & $g_{1L}$ &  Helicity & Convergent for $b\lesssim 2-3$GeV$^{-1}$.
\\\hline
$\gamma^+\gamma^5$ & $g_{1T}$ &  Worm-gear T & Convergent for $b\lesssim 2-3$GeV$^{-1}$. The convergence range  is possibly larger.
\\\hline
$\sigma^{\alpha+}\gamma^5$ & $h_{1}$ &  Transvercity & Convergent for $b\lesssim 4-5$GeV$^{-1}$.
\\\hline
$\sigma^{\alpha+}\gamma^5$ & \multicolumn{2}{|c||}{$h_{1L}^\perp$, $h_1^\perp$, $h_{1T}^\perp$} & Convergent in some large range of $b$.
\end{tabular}
\caption{\label{table} Convergence hierarchy for $\mathbf{M}^{(1)}_{\ln|x|}(b,\mu)$ evaluated on TMD PDFs of different kinds. The estimation of ranges for $b$ is made using values from \cite{Kovchegov:2016zex,Kovchegov:2018zeq,Scimemi:2019cmh}, and gives only the  general scale.}
\end{table*}

In Fig.\ref{fig:Mn} we plot $\mathbf{M}^{(n),unpol.}_{\ln|x|}$ as function of $b$ for different values of $n$ starting from $n=2$, using values from \cite{Scimemi:2019cmh}.  In Fig.\ref{fig:compare} we show the function $\mathbf{R}^{(2)}$ for typical lattice momenta $P_1^+=2.5$GeV and $P_2^+=2$GeV (and $|v^-|=1/\sqrt{2}$). The scale $\mu$ is set to be
$$\bar \mu=2|v^-|\sqrt{P_1^+P_2^+},$$
such that the logarithm in (\ref{def:rn}) is zero. As input we take two of the most recent extractions of unpolarized TMDPDFs and rapidity anomalous dimensions \cite{Scimemi:2019cmh,Bacchetta:2019sam}. The computation is done with the \texttt{artemide} package \cite{Scimemi:2017etj}. The uncertainty band is due to the uncertainty of the phenomenological parameters. The considered models have essentially different behavior at large values of $b$, which should be clearly distinguishable on the lattice. Let us also note that at small values of $b$,  the limit $R^{(n)}\to1$ is definitely violated. This is due to the $(|b|P^+)^{-1}$ correction that is part of $\mathcal{O}(\lambda)$.

To extract the rapidity anomalous dimension $\mathcal{D}$ from the lattice data, we have to deal with the function $\mathbf{M}_{\ln|x|}$. The value of $\mathbf{M}$ is difficult to estimate by lattice methods. However, we argue that the effects caused by nontrivial $\mathbf{M}$ can be neglected with reasonable accuracy. First of all, we mention that the corrections to $1$ in $\mathbf{r}^{(n)}$ have an extra suppressing factor $\ln(P_1^+/P_2^+)$. Together, with $\alpha_s/(4\pi)$ it guarantees that these corrections are numerically small. For example, for the parameter values used in Fig.\ref{fig:compare} the contribution of the term with $\mathbf{M}^{(2)}$ is $\sim 0.06$. Second, the function $\mathbf{M}$ has minor dependence on $b$. In Fig.\ref{fig:compare}, the contribution of maximum and minimum values of $\mathbf{M}^{(2)}$ (see Fig.\ref{fig:Mn}) differ by $\sim 0.008$, which is a tiny number in comparison to the expected accuracy of lattice computations. Based on this observation, we conclude that the function $\mathbf{M}^{(n)}$ could be replaced by a constant, adding a $\sim1\%$ systematic uncertainty. This constant can be estimated using the existing phenomenological extractions (with $\sim 1-2\%$ of systematic uncertainty), or it can simply be neglected (implying a $\sim 10\%$ systematic uncertainty). Therefore, the proposed method allows the determination of $\mathbf{R}^{(n)}$ within a few percent of systematic uncertainty, depending on the selected strategy. Alternatively, one can estimate (the constant) $\mathbf{M}^{(n)}$ from the lattice data at a single point.

Such simplified scheme should be applied with caution, because $\mathbf{M}$ has different behavior for different types of TMD distributions and different $n$. In the majority of cases, $\mathbf{M}_{\ln|x|}$ can be approximated by a constant (in $b$), however, in some cases not. It is clear that $\mathbf{M}_{\ln|x|}$ is closer to a constant if the integrand has better convergence properties at $x\to0$. However, for some cases the integrals in $\mathbf{M}_{\ln|x|}$ are divergent, such that these cases cannot be used for analysis. We should also keep in mind that higher perturbative terms contain $\mathbf{M}_{\ln^n|x|}$, and have worse convergence properties.

There are two sources of small-$x$ divergence in $\mathbf{M}$:

\textit{The first one} is the factor $|x|^{-2\mathcal{D}}$. The rapidity anomalous dimension $\mathcal{D}$ is greater then zero for $b\gtrsim 2e^{-\gamma_E}/\mu$. Its asymptotic behavior is unknown, although typically it is expected to be a monotonously growing function. Additionally, the value of $\mathcal{D}$ also increases with the increase of $\mu$. The uncertainty  of the large-$b$ behavior of modern extractions of $\mathcal{D}$ are quite drastic \cite{Bertone:2019nxa,Scimemi:2019cmh,Bacchetta:2019sam,Vladimirov:2020umg}. Nonetheless, all recent extractions agree that $\mathcal{D}>1/2$ for $b\gtrsim 3-4$GeV$^{-1}$ (here $\mu \sim 2$GeV). Therefore, in this range the factor $|x|^{-2\mathcal{D}}$ is singular.

\textit{The second one} is the TMD distribution itself. Generally, at small-$x$ TMD PDFs behave as $x^\alpha$. The value of $\alpha$ depends on the kind of TMD PDF, as has been studied in \cite{Kovchegov:2016zex,Kovchegov:2018zeq}. It has been shown (in the large-$N_c$ approximation) that $\alpha<-1$ for the unpolarized structure $\Gamma=\gamma^+$, $-1<\alpha<0$ for the helicity structure $\Gamma=\gamma^+\gamma^5$, and $\alpha>0$ for the transversity structure $\Gamma=\sigma^{+\mu}$. In each case only the leading distribution has been considered (i.e. $f_1$,$g_{1L}$ and $h_1$). One can expect weaker singularities with a similar general hierarchy for other distributions (i.e. $f_{1T}^\perp$, $g_{1T}$, etc). The power of the small-$x$ singularity also depends on the flavor combination. In particular, the nonsinglet combinations have weaker small-$x$ behavior (see e.g.\cite{Kovchegov:2016zex}).

In this way, there is a certain hierarchy of $\mathbf{R}$ for different $\Gamma$ and $n$, such that some give simpler access to $\mathcal{D}$ that other, see table \ref{table}. Note, that the convergence properties improve for nonsinglet flavor combination, and for larger $n$. In particular, the unpolarized case is convergent for $n=2$ in a large range of $b$, which is also seen in Fig.\ref{fig:Mn}.

Finally,let us stress again that the rapidity anomalous dimension is universal. Therefore, the ratios $\mathbf{R}^{(n)}$ should be almost independent of quark flavor, Dirac structure $\Gamma$, hadron type, and the momentum parameter $n$ (for convergent cases). The difference between all these cases is only due to functions $\mathbf{M}^{(n)}$.

\section{Conclusion}

In the present article, we have considered quasi-TMD operators, that can be investigated on the lattice. We pointed out the similarity of the lattice observable to the hadronic tensor of TMD processes, such as Drell-Yan or SIDIS. Using the method of soft-collinear effective field theory (SCET II) we derived the factorized expression for the lattice hadronic tensor in terms of physical TMD distributions, and the new instant-jet TMD distribution $\Psi$ defined in (\ref{def:Psi}), (\ref{Psi->physical}). The factorized expression generally coincides with expressions derived in \cite{Ebert:2019okf,Ji:2019ewn}, although the route of derivation is different. We have checked
factorization 
at one-loop level and derived the hard matching coefficient at this order, which coincides with the one derived in \cite{Ebert:2019okf}. The LO anomalous dimension can be extracted from the literature related to heavy-quark physics, and that value coincides with the results of our calculation. The present derivation is done for arbitrary Dirac structure, and can be easily extended to other interesting cases, such as gluon operators.

Since the factorization formula contains an unknown nonperturbative function $\Psi$, it is advantageous to consider the ratios of lattice observables with the same geometrical parameters of the operators (i.e. $\ell$, $b$, $L$ and $v$). In this case, many troublesome factors, such as $\Psi$ and lattice renormalization factors, cancel. The remaining parameters, namely the Dirac structure $\Gamma$, hadron momentum, spin and flavor, are enough to extract valuable information on TMD distributions and to estimate the uncertainties of the method. In particular, we pointed out that the ratio of the first derivatives at $\ell=0$ with different hadron momenta can be used to accurately determine the rapidity anomalous dimension (Collins-Soper kernel). In this case, one does not need to evaluate Fourier transformations with respect to $\ell$, as suggested in \cite{Ebert:2018gzl}. Evaluating  the ratios (\ref{R:suppresed}) for suppressed and unsuppressed Dirac structures  allows to estimate the systematic uncertainty of the method by lattice simulations. 

The hard scale of the derived factorization theorem is the hadron momentum $P$. Thus, one could expect that the corrections to the factorized term are $P^{-1}$-suppressed. However, this is only a crude estimate because the parton fields carry only a fraction of the total hadron momentum. Therefore, the true factorization scale is the parton momentum $xP$, which is generally much smaller. In contrast to the scattering processes, where the parton momentum is detected, lattice simulations involve all possible parton momenta. This leads to problems caused by low-x divergences. In particular, the power corrections to lattice factorizations are $1/x^2$-enhanced \cite{Braun:2018brg}. This observation limits the application range of such factorization approaches. In particular, in the $\ell=0$ case (that was considered in \cite{Musch:2011er,Engelhardt:2015xja}), the size of corrections is very strongly dependent on the operator. In certain cases (for instance unpolarized operators) already at NLO level one can encounter divergences. However, one is free to avoid such cases when determining $\mathcal{D}$.

Quite generally, many complications can be avoided if one considers the ratios of lattice observables. The information that could be extracted from ratios is limited, but still valuable. For instance, one can extract the nonperturbative rapidity anomalous dimension. The most simple observable in this case is the ratio of the first (and possibly the second) Mellin moments of quasi-TMDs at different hadron momenta. This ratio is almost exclusively dependent on $\mathcal{D}$. In addition, it depends on $\mathbf{M}^{(n)}$ defined in (\ref{def:M}). We argue that for most parts of combinations of quantum numbers $\mathbf{M}^{(n)}$ can be approximated by a constant with reasonable accuracy. The constant can either be estimated from phenomenology or by normalizing to the lattice data. In this way, one can determine the nonperturbative evolution kernel with a few percents of (theoretical) uncertainty.

\acknowledgements

Authors are thankful to V.~Braun, X.~Ji, Y.~Kovchegov, Y.~Liu, Y.-S.~Liu, M.~Schlemmer, I.~Stewart, and Y.-B.~Yang for stimulating discussions. This work was supported by DFG (FOR 2926 ``Next Generation pQCD for Hadron Structure: Preparing for the EIC'', project number 40824754).

\bibliography{TMD_lattice_ref}
\end{document}